\documentclass[aip,cha,reprint]{revtex4-2}

\usepackage[utf8]{inputenc}
\usepackage[T1]{fontenc}
\usepackage[british]{babel}
\usepackage{amsfonts,amssymb,amsmath}
\usepackage{booktabs}
\usepackage{longtable}
\usepackage{bm}
\usepackage{graphicx}
\usepackage{dcolumn}
\usepackage{textcomp}
\usepackage[colorlinks=true,linkcolor=blue,urlcolor=blue,citecolor=blue]{hyperref}
\usepackage{orcidlink}

\begin{document}

\title[\uppercase{\textsf{Pressure shifts in complex fluids under LAPS}}]{Pressure shifts in pulsatile shear: A microfluidic method to probe the normal stress response of complex fluids}

\date{November 2024}
\revised{December 2025}

\def\demec{CEFT, ALiCE, Dept.\ of Mechanical Engineering, Faculty of Engineering, University of Porto, Rua Dr.\ Roberto Frias, 4200-465 Porto, Portugal}
\def\deqb{CEFT, ALiCE, Dept.\ of Chemical and Biological Engineering, Faculty of Engineering, University of Porto, Rua Dr.\ Roberto Frias, 4200-465 Porto, Portugal}

\author{T.~\surname{Rodrigues}\orcidlink{0000-0003-1196-3796}}
\email[Corresponding author; electronic mail: ]{tomasrp@fe.up.pt}
\affiliation{\demec}

\author{F.~J.~\surname{Galindo-Rosales}\orcidlink{0000-0001-9763-6854}}
\affiliation{\deqb}

\author{L.~\surname{Campo-Dea{\~n}o}\orcidlink{0000-0003-2493-3238}}
\affiliation{\demec}

\begin{abstract}
	A microfluidic approach to probing the first normal stress difference from single-point pressure measurements in transient shear flows is presented. Using an original experimental design, we examine the near-zero-mean pulsatile flow of polymeric solutions in a straight microchannel at low Reynolds and Womersley numbers. An important aspect of this work is that the enhanced fluid elastic stresses can be efficiently determined via the pressure shift measured from pressure-controlled pulsatile shear experiments. We find a scaling law that collapses pressure-shift data from viscoelastic fluids of different molecular weights onto a single master curve that can then be used to predict this phenomenology. Taken together, these results could help shed light on our understanding of the non-linear normal stress responses in time-dependent flows.
\end{abstract}

\maketitle

\section{\label{sec:intro}Introduction}

Unlike water, complex fluids often have surprising behaviours due to their non-linear responses, including rod climbing, extrudate swelling, and flow instabilities. A particularly active research area is blood rheology (for a recent review, see Ref.~\onlinecite{beris2021}) and the influence of elastic microvessels on its dynamics. Blood is a complex biofluid with non-Newtonian characteristics,\cite{deano2013,sousa2013,rodrigues2022,brust2013} periodically pumped by the heart into a branching network of arteries. It is well established that non-linear effects manifest themselves in flows of viscoelastic liquids under large and rapid periodic forcing.\cite{saengow2020,sousa2013} Besides viscoelasticity, the fluid--structure interactions in microvessels differ significantly from those in large arteries.\cite{secomb2017} The vasodilatory capability of the former contributes to the decrease in shear and normal stresses that develop during flow.\cite{giannokostas2022} However, normal-stress effects are often overlooked, and some fundamental and practical issues remain open. Furthermore, understanding blood flow properties in microcirculation is an essential step towards elucidating health issues.\cite{thiebaud2014,rodrigues2020b,recktenwald2024}

Microfluidic platforms have proven to be very versatile for studies of low-viscosity complex fluids.\cite{qin2019,haward2012b,rosales2013,rodrigues2020b,recktenwald2024,brust2013,pipe2009} For example, the small confinement allows access to the high strain rates that, together with high strain amplitudes, define large-amplitude oscillatory shear~(\textsc{laos}) flow,\cite{saengow2020,ewoldt2016,sousa2013} with low inertia and tiny volume displacements. The use of micro-oscillatory flow to perform extensional rheometry or investigate polymer and vesicle dynamics has relied on different geometries.\cite{odell2006,zhou2016,bonacci2023,burgt2014,lin2021,recktenwald2025} However, generating precisely controlled oscillatory or \emph{pulsatile} shear flows---by superimposing oscillations onto a continuous flow---can be challenging.\cite{recktenwald2021,blythman2017,burgt2014}

Pressure measurements can be used to characterise the temporal structure of complex flows and rheological properties of polymeric fluids.\cite{yesilata2000} The measurement of normal force at the wall of a channel represents a combination of static pressure and primary normal stress due to fluid elasticity---often visible in extrudate swell tests.\cite{bird1987_book,walters1975_book,joseph1990_book} A convenient way of determining local normal stresses in non-Newtonian fluids is by attaching a pressure transducer at the bottom of a `small' hole in the wall.\footnote{Although more accurate alternatives\cite{gauthier2021,pipe2009} using flush-mounted transducers, albeit impractical in optically transparent microfluidic systems, have been proposed.} However, there is a difference between the pressure measured by the recessed transducer and the actual wall pressure. This measurement error was first confirmed by Lodge and co-workers (see, for example, Ref.~\onlinecite{broadbent1968}). The pressure difference has been termed the \emph{hole pressure} and is an effect mostly attributable to the relaxation of normal stresses in the streamlines.\cite{joseph1990_book,bird1987_book,yesilata2000,walters1975_book} In fact, \textcite{lodge1983} have suggested using it as a measure of the first normal stress difference~$N_1$.

In this paper, we present a microfluidic approach to probing $N_1$, a fundamental property of viscoelastic fluids, in pulsatile shear flows. Non-zero normal stresses, not seen in Newtonian fluids, generate elastic forces that are proportional to $N_1$, which can be indexed via the pressure shift from what we call `large-amplitude pulsatile shear'~(\textsc{laps} in short) microfluidic experiments. By defining a new dimensionless number, we obtain a scaling that collapses pressure-shift data from viscoelastic fluids having similar shear viscosity but different polymer molecular weight~$M_\text{w}$. The \textsc{laps} framework and the scaling law allow the quantification of the non-linear rheological response in complex fluids under industrially- and biologically-relevant flows, \textit{e.g.}, drilling, oil recovery, respiratory mucus flow, and blood pumping.

\section{\label{sec:methods}Materials and methods}

\subsection{\label{sec:methods:device}Microfluidic device}

\begin{figure}
	\includegraphics[width=\columnwidth]{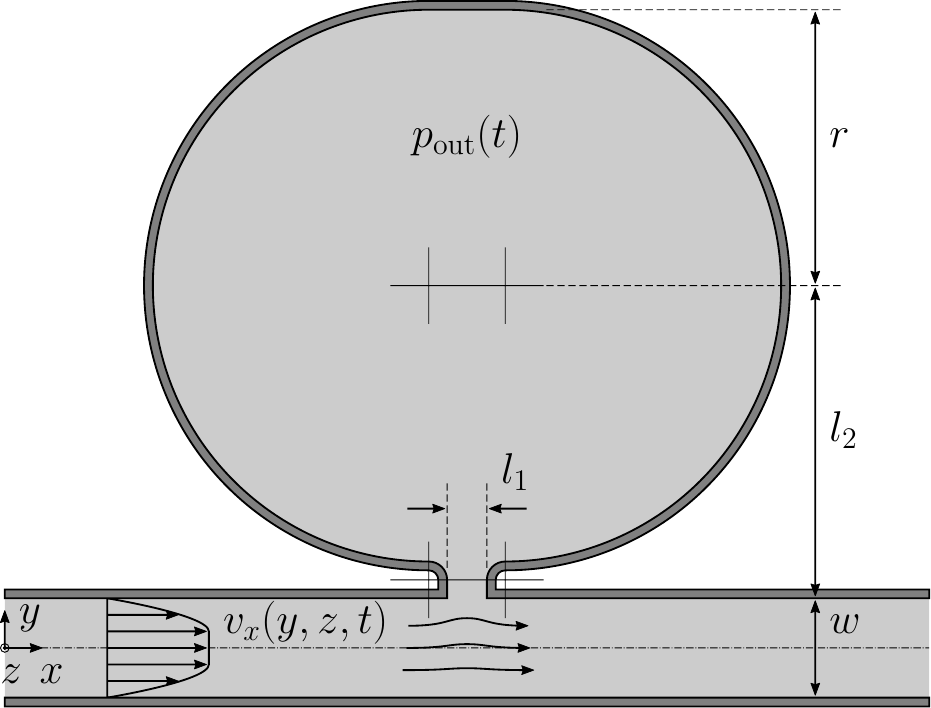}
	\caption{\label{fig:tap}Schematic of the $x$--$y$ plane of the microchannel and pressure tap~($p_\text{out}$) placed perpendicular to the flow direction ($r=750$~{\textmu m}, $l_2=850$~{\textmu m}). Net volumetric flow is left to right, driven by a sinusoidal pressure gradient. Flat, plug-like velocity profile~$v_x$ typical of shear-thinning polymer solution flows.}
\end{figure}

Experiments are performed in a long microchannel having a rectangular cross-section with depth $h=100$~{\textmu m} in the $z$ direction, width $w=270$~{\textmu m} in the $y$ direction and length 10~mm in the $x$ direction. The microchannel is fabricated out of polydimethylsiloxane~(\textsc{pdms}) using standard soft-lithography methods. Since the development length may be longer for polymeric solutions than Newtonian fluids,\cite{li2015} the pressure tap (Fig.~\ref{fig:tap}) is located 7.1~mm away from the inlet to avoid entrance effects and transport of any deformation history of the fluid into the measurement section. Furthermore, our smooth inlet (outlet) geometry likely shortens this region, and a relatively lengthy stretch ($>10w$) has been maintained downstream of the pressure tap. Therefore, these viscoelastic creeping flows, where the Reynolds~[$\text{Re}=\rho Q/(2w\eta_0)$] and Womersley~[$\text{Wo}=(h/2)\sqrt{\rho\omega/\eta_0}$]\cite{recktenwald2021,burgt2014,blythman2017,bonacci2023,lin2021} numbers are $O(10^{-3})$ or smaller (where $\rho$ is fluid density, $\eta_0$ is zero-shear viscosity, $Q$ is volumetric flow rate, and $\omega$ is angular frequency), are presumed to be fully developed at this location, and thus no correction is required.\cite{macosko1994_book} The size of the pressure slot is relatively small ($l_1=108$~{\textmu m}, $l_1/w=0.4$) to minimise disturbances to the flow such as hole-pressure effect.\cite{joseph1990_book,bird1987_book,yesilata2000,walters1975_book} Contributions from inertia to the hole pressure are negligible since the slot hydraulic diameter, $d_1=2l_1h/(l_1+h)$, is much smaller than the critical inertial diameter suggested by \textcite{joseph1990_book}, $d_\text{cr}=\sqrt{\lambda_\text{E}\eta_\text{p}/\rho}$ (with $\lambda_\text{E}$ and $\eta_\text{p}$ being the extensional relaxation time and polymer contribution to viscosity, respectively). Conversely, disturbances from normal stress dominate. Furthermore, we can neglect the effects of the second normal stress difference~$N_2$ on polymer solution flows ($N_2/N_1\ll 1$) across three-dimensional slots, following common practice in parallel-plate rheometry ($N_1-N_2\simeq N_1$). Based on the Tanner--Pipkin relation,\cite{joseph1990_book,bird1987_book,yesilata2000,walters1975_book} the hole-pressure error is estimated not to exceed 0.2\% of the absolute pressure measured.

\subsection{\label{sec:methods:samples}Sample solutions}

\begin{figure}
	\includegraphics[width=\columnwidth]{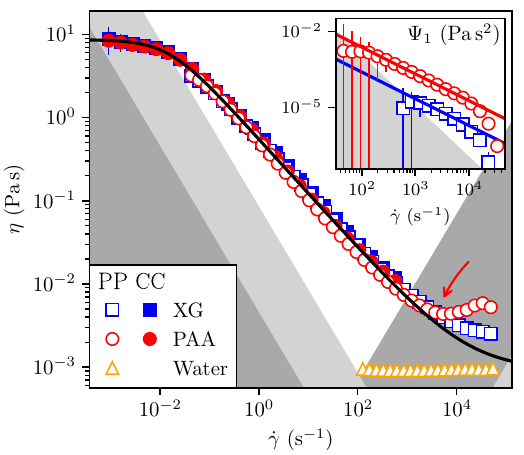}
	\caption{\label{fig:rheol}Fluid rheological characterisation. Shear viscosity~$\eta$ and first normal stress coefficient~$\Psi_1$ (determined where possible, inset plot) as a function of applied shear rate~$\dot\gamma$, with fits of \textsc{c--y} and power-law models, respectively. Minimum torque and secondary flow limits (slopes $-1$ and 1, respectively) shown in light and dark grey (\textsc{pp} and \textsc{cc} geometries, respectively).\cite{ewoldt2015} Inset:~Resolution limit (slope $-2$) shown in light grey.\cite{walters1975_book} The shear rate value reaches as high as $O(10^3~\text{s}^{-1})$ in the \textsc{laps} experiments.}
\end{figure}

Both Newtonian (water) and polymeric fluids are investigated. Two viscoelastic solutions with nearly identical shear-thinning viscosity but different longest polymer relaxation times are formulated: the first, referred to as \textsc{xg}, is a weakly elastic but strongly shear-thinning solution of 2500~ppm of xanthan gum dissolved in pure water; the second fluid, referred to as \textsc{paa}, is a viscoelastic and strongly shear-thinning solution made by adding 1000~ppm of polyacrylamide ($M_\text{w}=18$~MDa) to the same Newtonian solvent. The properties of the solutions discussed in this paper are summarised in Table~SI provided in the supplementary material. The fluids were characterised using a stress-controlled rheometer (Anton Paar MCR 301) with parallel-plate~(\textsc{pp}, $2R=50$~mm diameter, $H=0.1$~mm gap) and concentric-cylinder~(\textsc{cc}, cup diameter 29~mm, bob diameter 27~mm, bob length 40~mm) fixtures at 20~\textcelsius. The shear rheology of both the \textsc{xg} semi-rigid and \textsc{paa} flexible polymeric solutions is well-described by a single inelastic Carreau--Yasuda~(\textsc{c--y}) model\cite{bird1987_book} (Fig.~\ref{fig:rheol}). At $\lambda_\text{E}\dot\gamma\sqrt{H/R}\simeq 12$ in \textsc{pp} (where $H$ is the gap height and $R$ is the plate radius), an elastic instability\cite{ewoldt2015} is observed for \textsc{paa} (arrow in Fig.~\ref{fig:rheol}). Similar instabilities in pressure-driven microchannel flow have been reported.\cite{qin2019} With increasing $\dot\gamma$, the first normal stress coefficient~$\Psi_1$ is described by the power-law scaling~$\Psi_1=b\dot\gamma^{m-2}$, with $m\simeq 0.9$ (inset of Fig.~\ref{fig:rheol}). Like for $\Psi_1$ ($\Psi_1^\text{PAA}\simeq 10\Psi_1^\text{XG}$), the measured $\lambda_\text{E}$ (using capillary breakup extensional rheometry) of the \textsc{paa} fluid ($\simeq 72.4$~ms) is about an order of magnitude larger than that of the \textsc{xg} fluid ($\simeq 8.2$~ms) (see the supplementary material, Fig.~S1). The storage modulus~$G'$ measured in small-amplitude oscillation---output linearly dependent on input---is slightly higher for the \textsc{xg} fluid (see Fig.~S2 in the supplementary material), typical of semi-rigid polymeric solutions in the limit of small deformations.

\subsection{\label{sec:methods:setup}Experimental setup}

The planar pressure-driven flow is imposed through a pressure controller with a stability and a precision of approximately 0.01~kPa and a typical response time of 50~ms (Elveflow AF1 Dual), connected to a flow sensor with an uncertainty of 5\% and a response time of 40~ms (Elveflow MFS4). This allows precise control of the applied pressure and flow rate measurement. The desired pressure is supplied to the cap of a pressure vessel partially filled with the working fluid. Flow exits the microfluidic channel at atmospheric pressure. Gauge pressures are measured with a piezoresistive, diaphragm-type sensor capable of resolving pressure differences of about 0.03~kPa (Silicon Microstructures SM5852 series, accuracy $\pm 1.6\%$ full scale, range 0--2.1~kPa), operating at 30~Hz. Details of the pressure sensor calibration are provided in the supplementary material, in Fig.~S3. The sensor is connected via flexible Tygon tubing to a very small hole (0.5~mm in diameter) punched through the \textsc{pdms} as an access port for the pressure tap~($p_\text{out}$). Further details of the pressure measurement validation can be found in \textcite{rodrigues2020b}. Voltage is read by a data acquisition card (USB-6218, National Instruments) working with a custom Lab\textsc{view} program. A global trigger signal is used to synchronise the controller and pressure sensor. The response time of the pressure measuring system depends on the deformability of the channel and pressure tap, elasticity and length of the tubing, pressure slot size, compliant air in the sample container and pressure port, and fluid properties.\cite{recktenwald2021,yesilata2000,burgt2014} To avoid microscopic air bubbles trapped in the pressure tap, care is taken during bleeding using the gas permeability of the \textsc{pdms} channel walls prior to experiments. Young's modulus for \textsc{pdms} is in the range of 0.5--4~MPa (mostly depends on curing conditions) and the \emph{applied} system pressure marginally exceeds 50~kPa for the worst cases, so the (soft) top wall of the channel is unlikely to deform significantly for a depth to width ratio of $h/w\simeq 0.4$ (the substrate is glass),\cite{gervais2006} especially when considering that the \textsc{pdms} slab (several millimetres) is much thicker than the width of the channel.\cite{hardy2009} The system's compliance is reduced to a minimum by using rigid polytetrafluoroethylene~(\textsc{ptfe}) and stiff polyetheretherketone~(\textsc{peek}) tubing connections between the sample reservoir, the flow sensor, and the microfluidic device, in part due to the shear-thinning nature of the viscoelastic fluids. The Tygon, \textsc{ptfe}, and \textsc{peek} tubing are kept as short as possible.\footnote{The Tygon tubing could not be eliminated altogether since the port on the pressure sensor is best sealed by flexible tubing.} All experiments are performed at room temperature.

\section{\label{sec:results}Results}

\begin{figure}
	\includegraphics[width=\columnwidth]{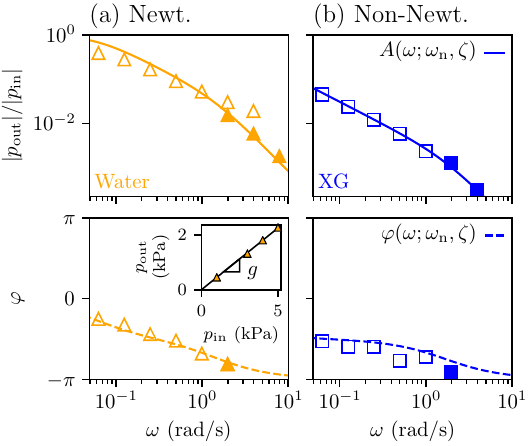}
	\caption{\label{fig:bode}Frequency response of the pressure measuring system. Bode diagrams of experimental data for the transfer functions of the (a)~Newtonian ($\omega_\text{n}=0.29$~rad/s, $\zeta=2.63$) and (b)~non-Newtonian ($\omega_\text{n}=0.07$~rad/s, $\zeta=11.77$) second-order systems, with overdamped dynamics ($\zeta>1$) reminiscent of that of soft materials. $\omega_\text{n}$ and $\zeta$ are the natural frequency and damping ratio of the system, respectively, which are typically present in the transfer function of a second-order system. The non-Newtonian data correspond to $10\agt\text{Wi}_\text{max}\agt 7$ and $8.2\times 10^{-5}\alt\text{De}\alt 1.3\times 10^{-3}$ (hollow squares); $\text{Wi}_\text{max}\simeq\{26,7\}$ and $\text{De}\simeq\{2.6,5.2\}\times 10^{-3}$ (filled squares). Different symbols represent different input mean-amplitude pairs $(\langle p_\text{in}\rangle,|p_\text{in}|)$ (see also Fig.~S4 in the supplementary material). The inset in (a) shows $p_\text{out}$ \textit{versus} $p_\text{in}$, used to determine $g=0.45$. Error bars (based on pressure sensor resolution) are less than marker size and are not shown here for clarity. Colour code as in Fig.~\ref{fig:rheol}.}
\end{figure}

The present experimental measurements involve pressure wave propagation in viscoelastic materials in a three-dimensional domain with non-rigid walls.\cite{yesilata2000} The parameters influencing the fluid pressure response are determined and modelled to overcome the experimental difficulties associated with generating sinusoidally-pulsating flows. \textcite{recktenwald2021} decomposed hydrodynamic parameters into frequency- and system-dependent amplitude attenuation~$A$ and phase shift~$\varphi$ relative to a prescribed pressure input~$p_\text{in}=\langle p_\text{in}\rangle+|p_\text{in}|\sin(\omega t)$. In the context of control theory, first one measures experimentally the system \emph{transfer function} in the frequency domain on the basis of the response to sinusoidal inputs (Fig.~S4 in the supplementary material); second one fits a model transfer function in polar form [Eqs.~(4) and (5) of Ref.~\onlinecite{recktenwald2021}] to the Bode amplitude and phase plots of frequency-response data (Fig.~\ref{fig:bode}). These aspects are discussed in detail elsewhere.\cite{recktenwald2021,burgt2014} Additionally, one must consider a second gain~$g$ that accounts for pressure head loss due to viscous stresses between the driving signal~$p_\text{in}$ generated by the flow controller and the actual output pressure~$p_\text{out}$ measured \textit{in situ} [determined herein as their steady-state ratio in the Newtonian solvent limit~($\dot\gamma\to\infty$), see the inset of Fig.~\ref{fig:bode}(a)]. We obtain the time-dependent linear response function
\begin{equation}
	\begin{array}{cccccl}
		p_\text{out}(t) & = & g\langle p_\text{in}\rangle & + & A(\omega)|p_\text{in}| & \sin[\omega t+\varphi(\omega)] \\
		 & \equiv & p_\text{off} & + & p_0 & \sin(\omega t)
		\,,
	\end{array}
	\label{eq:3p}
\end{equation}
which is reversed to `calculate an adapted, optimized system input'\cite[{p.~2608}]{recktenwald2021} on a case-by-case basis. The applied system pressure now \emph{compensates} for (i)~frictional losses, (ii)~damping, and (iii)~phase lag, allowing for precise control of the instantaneous pressure in the parallel shear flow section (Fig.~\ref{fig:tap}).

\begin{figure}
	\includegraphics[width=.7\columnwidth]{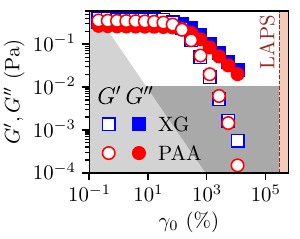}
	\caption{\label{fig:laos}Strain amplitude sweep for \textsc{xg} and \textsc{paa} at a fixed frequency~$\omega=0.79$~rad/s. Beyond the linear regime, both $G'$ and $G''$ monotonically decrease. The non-linear \textsc{laps} region is marked as the pink area. The minimum torque and instrument inertia limits are shown in light and dark grey, respectively.\cite{ewoldt2015} Error bars are less than marker size and are not shown here for clarity. Colour code as in Fig.~\ref{fig:rheol}.}
\end{figure}

A transfer function is a measure of the response of a system to a given input. Generally, the experimental (or measured) transfer function, and hence the fit parameters $(\omega_\text{n},\zeta,g)$, change when the mobile fluid system is changed (visualised in Fig.~\ref{fig:bode}). Since both non-Newtonian fluids share nearly the same viscous response (Fig.~\ref{fig:rheol}) and the characteristic time of the driving~$T=2\pi/\omega$ is longer than their relaxation time~$\lambda_\text{E}$, \textit{i.e.}, the Deborah number~($\text{De}=\lambda_\text{E}/T$) is smaller than unity, determining the frequency response using one or the other should result in nearly identical transfer functions. This is consistent with the close agreement between steady-state flow rate measurements taken for a range of applied system pressures with the \textsc{xg} and \textsc{paa} test fluids (see the supplementary material, Fig.~S5). However, we note that in practice, if the parameters vary from fluid to fluid---or from, say, setup to setup---the system's frequency response may need to be optimised and recalibrated between samples.

\begin{figure}
	\includegraphics[width=.8\columnwidth]{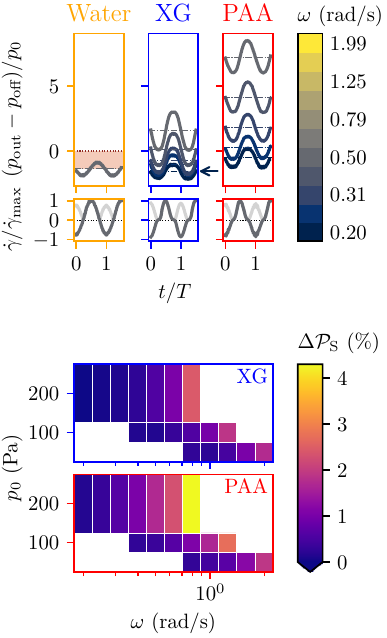}
	\caption{\label{fig:laps}Pressure shift evolution. Top:~Frequency sweep responses at $p_0=200$~Pa (left column for water, middle column for \textsc{xg}, and right column for \textsc{paa}). The non-Newtonian data show a vertical shift, indicating the presence of viscoelasticity. This effect becomes more pronounced for increasing $\omega$, and the corresponding pressure shifts are labelled as dash-dotted lines. This phenomenology was subsequently reproduced using a blood-mimicking fluid and a physiological waveform (this will be discussed in a future paper), suggesting that our results should apply more widely. Bottom:~Magnitude of pressure shift~$\Delta\mathcal{P}_\text{S}$ as a function of imposed frequency~$\omega$ and pressure amplitude~$p_0$. In the white region, there are no data.}
\end{figure}

A set of \textsc{laps} experiments sweeps both the angular frequency~$\omega$ and pressure amplitude~$p_0\equiv A|p_\text{in}|$. Provided that the Weissenberg number~[$\text{Wi}=\lambda_\text{E}\dot\gamma$, evaluated at the wall shear rate~$\dot\gamma=2Q/(wh^2)$] and the dimensionless strain~$\text{Wi}/\text{De}$ are sufficiently large,\cite{ewoldt2016} a \textsc{laps} experiment accesses non-linear viscoelasticity, as confirmed by comparing the \textsc{laps} region (where $2992\alt\text{Wi}_\text{max}/\text{De}\alt 42\,121$, calculated based on the maximal shear rate) to strain-sweep data (Fig.~\ref{fig:laos}). Owing to the large shear rates present, we explore a broad range of Pipkin space spanning $3\alt\text{Wi}_\text{max}\alt 297$ and $2.6\times 10^{-4}\alt\text{De}\alt 2.3\times 10^{-2}$, depending on the fluid under investigation. For completeness, we show such a plot in the supplementary material, in Fig.~S6. Representative \textsc{laps} frequency sweeps are shown in Fig.~\ref{fig:laps}~(top). We observe the periodic modulation of the \textit{in situ} pressure measurement about a mean steady-state value. For the polymeric solutions, the resulting shear rate waveforms are distorted from sinusoidal waves due to the very broad shear-thinning response noticeable in the `core regions' of high shear rate. Additionally, we observe a phase shift between the shear rate (bottom row) and pressure (top row) signals, independent of fluid rheology and applied pressure. It indicates that the output pressure lags behind the flow. Since in all cases $\text{Wo}\ll 1$,\cite{recktenwald2021,bonacci2023} this time delay is likely due to the air in the pressure port or compliance of the pressure tubing. For the Newtonian control case and all flow conditions $(\omega,p_0)$, time-averaged pressure data agree with predictions of Eq.~(\ref{eq:3p}) to within 0.4\% of the absolute value. The highlighted region in Fig.~\ref{fig:laps}~(top) indicates that the Newtonian data are close to zero, but not identically zero. We use this as a baseline measurement for comparison of the non-Newtonian results by subtracting the constant offset, which deviated from zero within experimental uncertainty. At smaller maximum shear rates~$\dot\gamma_\text{max}$, or flow strengths~$\text{Wi}_\text{max}$, these results are also broadly consistent with the predicted (baseline) response, similar to the Newtonian case, as indicated by the arrow in Fig.~\ref{fig:laps}~(top) for $\text{Wi}_\text{max}^\text{XG}\simeq 4$. We might call this the `moderately' non-linear (viscous-dominated) region of the $(\omega,p_0)$ parameter space (the lower left). At higher $\omega$ and/or $p_0$ (note the functional dependence of the inverse gain~$1/A$ on $\omega$), the onset of elastic effects is closely associated with a vertical shift between the measured and predicted pressure waveforms. In other words, the pressure
\begin{equation}
	p_\text{out}(t)\simeq p_\text{off}+\underbrace{p_0\sin(\omega t)+\Delta p_\text{S}(\omega,p_0)}_\text{$p_\text{out}(t)-p_\text{off}$}
	\label{eq:4p}
\end{equation}
oscillates about an offset~$p_\text{off}+\Delta p_\text{S}$ rather than just the first term as in Eq.~(\ref{eq:3p}). After baseline subtraction, the vertical displacement term~$\Delta p_\text{S}$ defined by the pressure-shift approach accounts for elastic stresses and reduces to zero in the moderately non-linear regime where these can be neglected. It is thus an index that describes in some way a degree of non-linearity. Departures from moderately non-linear responses increase with Wi---and so does $N_1$---accompanied by an amplification of the modulation component ($|p_\text{out}|\agt p_0$).

\begin{figure}
	\includegraphics[width=.8\columnwidth]{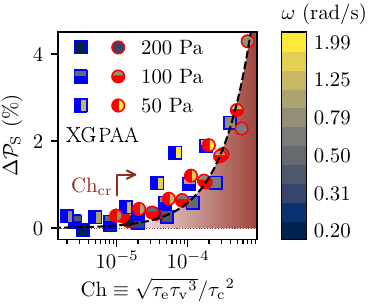}
	\caption{\label{fig:chronos}Master curve of the normalised pressure shift~$\Delta\mathcal{P}_\text{S}$ \textit{versus} the dimensionless combination~$\sqrt{\tau_\text{e}{\tau_\text{v}}^3}/{\tau_\text{c}}^2$~($\equiv\text{Ch}$). The dotted line indicates 0\% change from the baseline. Above a critical Chronos number ($\text{Ch}_\text{cr}\simeq 10^{-5}$), where the data points deviate from this baseline response, shifting occurs. The filled and half-filled symbols are coloured based on their corresponding angular pulsation frequencies (face colours represent $\omega$, as indicated by the colour bar). The dashed line shows a linear scaling as a guide to the eye.}
\end{figure}

We define the magnitude of elastic normal stresses via the normalised pressure shift~$\Delta\mathcal{P}_\text{S}=\Delta p_\text{S}/p_\text{off}$, where $p_\text{off}$ is the time-averaged pressure in the absence of elastic effects. Values of this ratio near zero indicate negligible shear normal stresses. To summarise the regimes of $(\omega,p_0)$ in which this effect arises, we show in Fig.~\ref{fig:laps}~(bottom) dynamic phase diagrams where each coordinate pair corresponds to a \textsc{laps} experiment with those given $(\omega,p_0)$. Represented by the coloured block at each $(\omega,p_0)$ is the magnitude of pressure shifting~$\Delta\mathcal{P}_\text{S}$ that arises in that particular experiment. Three frequency sweeps (at different fixed amplitudes) are used to create each parameter space `fingerprint' shown in Fig.~\ref{fig:laps}~(bottom). The ability to separately vary the amplitude and frequency of the imposed pressure provides a rheological fingerprint in a phase diagram that characterises the non-linear normal stress response. The shear flow reverses direction and, hence, momentarily vanishes twice during a \textsc{laps} cycle; the instantaneous shear rate is not uniformly large. Thus, it is helpful to restrict our analysis to $\text{Wi}=\text{Wi}_\text{max}$, where the polymer molecules are strongly perturbed from equilibrium. The wall shear rate takes on its maximal value, $\dot\gamma=\dot\gamma_\text{max}$, at this point in time. The \textsc{laps} response at high Wi results from a competition between the elastic~$\tau_\text{e}=\lambda_\text{E}$, viscous~$\tau_\text{v}=L^2\rho/\eta_0$, and convective~$\tau_\text{c}=1/\dot\gamma_\text{max}$ timescales, where $L=h/2$ is channel half-depth. These characteristic measures can be used to arrive at some degree of collapse of the pressure-shift data in Fig.~\ref{fig:laps}~(bottom) by plotting $\Delta\mathcal{P}_\text{S}$ against the dimensionless quantity~$\sqrt{\tau_\text{e}{\tau_\text{v}}^3}/{\tau_\text{c}}^2$. In general, we postulate the normalised pressure shift to scale as
\begin{equation}
	\Delta\mathcal{P}_\text{S}\propto\frac{\sqrt{\tau_\text{e}{\tau_\text{v}}^3}}{{\tau_\text{c}}^2}
	\,,
	\label{eq:chronos}
\end{equation}
which is in fair agreement with our experimental data, as illustrated in Fig.~\ref{fig:chronos}. We, thus, propose to define a \emph{Chronos number}~(Ch) as the combination of three competing timescales as per Eq.~(\ref{eq:chronos}), which is equivalent to the product of the $\text{Re}\equiv\tau_\text{v}/\tau_\text{c}$ and \emph{viscoelastic} Mach~($\text{Ma}_\text{ve}=\sqrt{\text{Re\,Wi}}\equiv\sqrt{\tau_\text{e}\tau_\text{v}}/\tau_\text{c}$)\cite{rodrigues2020b} numbers. From an empirical perspective, Ch captures the main variations of $\Delta\mathcal{P}_\text{S}$ with these particular experimental parameters. The quantity defined in Eq.~(\ref{eq:chronos}) is pertinent to the present data because it combines elasticity, viscosity, and inertia in a manner where $\text{Ch}\agt 10^{-5}$ is expected to generate appreciable normal stresses. We also find that Ch and the dimensionless groupings~$\mathcal{X}=\{\text{Re},\text{Wi},\text{Ma}_\text{ve}\}$ satisfy a relationship with the elasticity number~($\text{El}\equiv\tau_\text{e}/\tau_\text{v}$),\cite{rodrigues2020b,rosales2013,li2015} $\text{Ch}/\mathcal{X}^2=\text{El}^\kappa$ with $\kappa=\{1/2,-3/2,-1/2\}$. The strong correlation between $\Delta\mathcal{P}_\text{S}$ and Ch based on Eq.~(\ref{eq:chronos}) shows that this dimensionless number should be useful to rheologists for quantifying deviations from the \emph{moderately non-linear limit} ($\text{Ch}>\text{Ch}_\text{cr}$), indicating that fluid elastic effects are non-negligible.

\begin{figure}
	\includegraphics[width=\columnwidth]{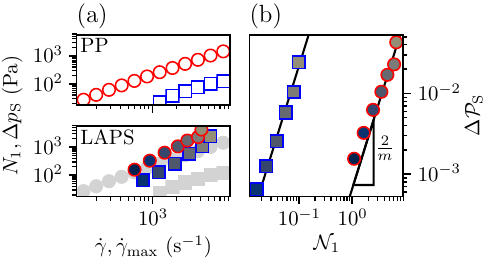}
	\caption{\label{fig:n1}Correlation to $N_1$. (a)~Comparison of the pressure-shift (\textsc{laps}) data to normal-stress data~$N_1$ obtained from a parallel-plate~(\textsc{pp}) device for \textsc{xg} and \textsc{paa} (squares and circles, respectively). (b)~The normalised pressure shift~$\Delta\mathcal{P}_\text{S}$ is plotted as a function of dimensionless first normal stress difference~$\mathcal{N}_1$ for both polymeric solutions, where the normal-stress dependence of $\Delta\mathcal{P}_\text{S}\propto{\mathcal{N}_1}^{2/m}$ is observed. Symbols as in Figs.~\ref{fig:rheol} and \ref{fig:chronos}. In the bottom panel of (a), the symbols in light grey represent the $N_1$ data.}
\end{figure}

As already discussed in Sec.~\ref{sec:intro}, pressure measurement is an essential component for performing (microfluidic) rheometry in channels and is a highly desirable metric in all flow studies.\cite{pipe2009,gauthier2021} Our primary interest here is to demonstrate the potential of a new microfluidic method in the measurement of $N_1$ from pressure-shift data. We present in Fig.~\ref{fig:n1}(a), a comparison of the \textsc{laps} data to $N_1$ data. We observe that the \textsc{laps} method slightly over-predicts the actual measured first normal stress difference. The normal stress difference measured by the axial force transducer of the rheometer is normalised by the zero-shear-rate elastic modulus~$G_0=\eta_0/\lambda_\text{E}$. The normalised pressure shift is plotted as a function of $\mathcal{N}_1=N_1/G_0$ in Fig.~\ref{fig:n1}(b). The data arrange on two distinct lines. Both the investigated polymeric solutions show that the pressure shift scales with the normal stress difference as $\Delta\mathcal{P}_\text{S}\propto\text{Ch}\propto{\mathcal{N}_1}^{2/m}$, where the value of $2/m$ is set by the degree of shear-thinning in the elastic normal stresses. By definition, dividing $\mathcal{N}_1$ by $\text{Wi}^2$ gives the dimensionless first normal stress coefficient~$\breve\Psi_1=\Psi_1/(G_0{\lambda_\text{E}}^2)$, therefore $\Delta\mathcal{P}_\text{S}\propto\text{Ch}\propto{\breve\Psi_1}^{2/(m-2)}$, where $m$ is the same as above. In non-linear viscoelasticity, the (extra) pressure shift contains information relating to the first normal stress difference, which can be useful in studying polymeric solutions. Overall, we have demonstrated that the pressure shift can be used as a \emph{rheological indexer} to quantify the normal-stress behaviour of low-viscosity complex fluids undergoing high-rate deformations in microfluidic devices.

We note that \textcite{zell2010} have proposed an empiricism for relating $\lambda_\text{E}$ to normal-stress data in the limit of zero shear rate, but this relation is inconsistent with our scaling. The results reported therein are for solutions exhibiting weak shear-thinning and quadratic normal stresses ($N_1\propto\dot\gamma^2$), for which $\Psi_1$ is constant by definition. This absence of shear-thinning effects in the viscometric functions is in stark contrast to the strongly (non-linear) shear-thinning behaviour of our fluids, since, for the shear rates at which $N_1$ was measured, the viscosity of neither fluid is close to the zero-shear-rate value. Their analysis, hence, ignores the key shear-thinning physics that dominates at sufficiently large $\text{Wi}_\text{max}$, most notably the shear-thinning in the elastic normal stresses within the range of shear rates obtained here ($377~\text{s}^{-1}\alt\dot\gamma_\text{max}\alt 5303~\text{s}^{-1}$). Lastly, we would like to bring the attention towards the disparity between relaxation times measured in shear and extensional flow, which precludes a direct comparison between these results and those reported in their paper.

\section{\label{sec:summary}Concluding remarks}

In summary, the distinct phenomenology of pressure responses being vertically shifted from the predicted waveform [the offset term in Eq.~(\ref{eq:3p})] under pressure-driven pulsatile shearing has been investigated. We have argued that the additional pressure shift in the non-periodic part of the response comes from an `elastic' contribution. The experimental design in this paper focused on isolating the effect of normal stresses resulting from very large straining motions. To that end, we have introduced the pressure-controlled \textsc{laps} framework. The \textsc{laps} method provides a physical interpretation of deviations from what we called \emph{moderately} non-linear behaviour, decomposing viscous and elastic contributions additively such that $p=p_\text{v}+\Delta p_\text{e}$ as in Eq.~(\ref{eq:4p}). The magnitude of the additional pressure originating from polymer elasticity ($\Delta p_\text{e}$) is an indication of non-linear viscoelastic effects, suggesting that it is an index of $N_1$. By defining a timescale-based parameter, denoted Ch, in terms of two extant dimensionless groups $(\text{Re},\text{Ma}_\text{ve})$, we obtained a scaling relation [Eq.~(\ref{eq:chronos})] that collapses pressure-shift data from \textsc{laps} measurements made with viscoelastic fluids having approximately similar $\eta_\text{p}$ but different elastic properties.

Transient \textsc{laps} microfluidic experiments may prove to be, in some cases, more convenient in detecting very small normal stresses down to very small polymer concentrations than steady shear. The technique's key benefits are the even smaller volume of sample required ($<1.5$~ml) and the ability to deliver high shear rates at relatively low Re. Additionally, due to the low solute concentrations and weak viscoelasticity, measuring the normal stress under shear flow for blood, plasma, synovial fluid, saliva, mucus, and vitreous humour, to name a few examples, is typically more difficult using classical rheometry techniques.\cite{beris2021,rosales2013,pipe2009} In principle, \textsc{laps} tests can be used to characterise a wide range of complex biological fluids under extremely non-linear flow conditions typical in physiological flows.

The present framework and the relationship allow the quantification of the moderate-to-strongly non-linear rheological response of complex fluids under pulsatile conditions. These can be useful to rheologists for industrial, academic, and biomedical research purposes. Our framework raises manifold avenues for efficiently probing non-linear elastic effects at high Wi but low Re or Wo. It would be interesting in future work to test the limits of the proposed scaling by widening the range of parameters under study, starting with fluid rheology as it governs many biological processes.

\section*{\label{sec:supplement}Supplementary material}

See the supplementary material for further supporting data and details on rheology and experimental methods.

\begin{acknowledgments}
	T.~R.\ thanks Professor~F.~T.~Pinho for helpful comments on an earlier draft of the manuscript.
	We thank the anonymous reviewers for their constructive criticisms.
	This work was supported by FCT ({2021.\allowbreak 06532.\allowbreak BD}), with additional support by FCT/MECI ({UID/\allowbreak 00532/\allowbreak 2025}, {UID/\allowbreak PRR/\allowbreak 00532/\allowbreak 2025}, {LA/\allowbreak P/\allowbreak 0045/\allowbreak 2020}, {2020.\allowbreak 03203.\allowbreak CEECIND/\allowbreak CP1590/\allowbreak CT0005}, and {PTDC/\allowbreak EME-APL/\allowbreak 3805/\allowbreak 2021}).
\end{acknowledgments}

%

\end{document}